\documentclass[amsmath,amssymb,pra,aps,showpacs,superscriptaddress,twocolumn]{revtex4-1}
\usepackage{amsmath,amsfonts,amssymb,amsthm,graphics,graphicx,epsfig,bbm}
\usepackage[colorlinks=true,citecolor=blue,linkcolor=blue,urlcolor=blue]{hyperref}
\usepackage[usenames]{color}
\usepackage{graphicx}
\usepackage{subfigure}
\usepackage{amsmath}
\usepackage{epsfig}
\usepackage{dcolumn}
\usepackage{bm}
\usepackage{color}
\usepackage{times}
\usepackage{epstopdf}
\usepackage{amssymb}
\usepackage{amstext}
\usepackage{latexsym}
\usepackage{hyperref}
\usepackage{amsfonts}
\usepackage{psfrag}
\usepackage{soul,xcolor}
\usepackage[normalem]{ulem}
\usepackage{dsfont}
\usepackage{txfonts}

\newcommand{\ket}[1]{\vert #1 \rangle}
\newcommand{\bra}[1]{\langle #1 \vert}

\begin{document}
\setstcolor{red}
	
\title{One-way quantum computing in superconducting circuits} 
\date{\today}
	
\author{F. Albarr\'an-Arriagada}
\email[F. Albarr\'an-Arriagada]{\qquad francisco.albarran@usach.cl}
\affiliation{Departamento de F\'isica, Universidad de Santiago de Chile (USACH), Avenida Ecuador 3493, 9170124, Santiago, Chile}
	 
\author{G. Alvarado Barrios}
\affiliation{Departamento de F\'isica, Universidad de Santiago de Chile (USACH), Avenida Ecuador 3493, 9170124, Santiago, Chile}
	
\author{M. Sanz}
\affiliation{Department of Physical Chemistry, University of the Basque Country UPV/EHU, Apartado 644, 48080 Bilbao, Spain}
	
\author{G. Romero}
\affiliation{Departamento de F\'isica, Universidad de Santiago de Chile (USACH), Avenida Ecuador 3493, 9170124, Santiago, Chile}
	
\author{L. Lamata}
\affiliation{Department of Physical Chemistry, University of the Basque Country UPV/EHU, Apartado 644, 48080 Bilbao, Spain}
	
\author{J. C. Retamal}
\affiliation{Departamento de F\'isica, Universidad de Santiago de Chile (USACH), Avenida Ecuador 3493, 9170124, Santiago, Chile}
\affiliation{Center for the Development of Nanoscience and Nanotechnology 9170124, Estaci\'on Central, Santiago, Chile}
	
\author{E. Solano}
\affiliation{Department of Physical Chemistry, University of the Basque Country UPV/EHU, Apartado 644, 48080 Bilbao, Spain}
\affiliation{IKERBASQUE, Basque Foundation for Science, Maria Diaz de Haro 3, 48013 Bilbao, Spain}
\affiliation{Department of Physics, Shanghai University, 200444 Shanghai, China}
	
\begin{abstract}
We propose a method for the implementation of one-way quantum computing in superconducting circuits. Measurement-based quantum computing is a universal quantum computation paradigm in which an initial cluster-state provides the quantum resource, while the iteration of sequential measurements and local rotations encodes the quantum algorithm. Up to now, technical constraints have limited a scalable approach to this quantum computing alternative. The initial cluster state can be generated with available controlled-phase gates, while the quantum algorithm makes use of high-fidelity readout and coherent feedforward. With current technology, we estimate that quantum algorithms with above 20 qubits may be implemented in the path towards quantum supremacy. Moreover, we propose an alternative initial state with properties of maximal persistence and maximal connectedness, reducing the required resources of one-way quantum computing protocols.

\end{abstract}
	
\maketitle

\section{Introduction}
Quantum computation has experienced a fast and remarkable development  in the last decades \cite{Taylor2005,Kok2007,Raussendorf2007,Haffner2008,Ladd2010,Saffman2010,Barends2014,Barends2016,Bernien2017,Lekitsch2017}. This progress is based on the astonishing development of quantum platforms  that currently allows for the manipulation and control of highly coherent qubits. There are several equivalent quantum computing paradigms for the implementation of a quantum algorithm. The quantum circuit or gate-based approach, which makes use of single- and two-qubit gates for this implementation is currently the standard one. One-way quantum computing, also known as measurement-based quantum computation \cite{Raussendorf2001,Raussendorf2003,Raussendorf2006}, is an alternative universal quantum computing paradigm, in which a particular kind of entangled multiparticle state, namely the cluster state, constitutes the initial quantum resource. Then, the algorithm is encoded through a sequence of single-qubit readouts and feedforward rotations based on the outcome of these measurements. Cluster states show two relevant properties, namely, persistence of entanglement after single-qubit projective measurements and maximal pairwise connectedness between qubits of the multipartite system. The first property refers to the minimal number of qubits that must be measured such that the resulting state is separable. Cluster states have a persistence of $\lfloor N/2 \rfloor$ \cite{Eisert2001}. The second property means that any pair of qubits can be projected onto a Bell state by appropriate local measurements on the rest of the qubits.
	
One-way quantum computing has been experimentally demonstrated in quantum photonics for small systems \cite{Walther2005experimental,Prevedel2007high,Tame2007experimental}. However, these implementations are not scalable due to the non-deterministic generation of the cluster state. Nonetheless, there have been theoretical proposals to avoid these issues \cite{Wang2010}. On the other hand, the proposals in trapped ions \cite{Stock2009scalable,Lanyon2013measurement} allow for scalable generation of cluster states, but the implementation of quantum algorithms is limited by the long qubit readout times. Indeed, measurements based on electron shelving ($\sim$$10$~ms \cite{Leibfried2003quantum}) take longer than the average  coherence time of the ions ($\sim$$3$~ms \cite{Gerritsma2011quantum}), which is a handicap for large protocols.

There has been remarkable progress in technologies based on superconducting circuits in recent years \cite{Devoret2004,Blais2007,You2011,Bulata2011,Gu2017}, which have allowed the scalable generation of large entangled states \cite{Neeley2010,Barends2016,Song201710qubits}. Additionally, fast feedforward protocols have also been developed recently \cite{Riste2012A,Riste2012D,Vijay2012}. Although theoretical proposals for the generation of cluster states in superconducting circuits have been made over a decade ago \cite{Tanamoto2006,You2007,Tanamoto2009,Wu2010}, the state of the technology at the time made difficult the implementation of these approaches.
		
In this work, we propose an experimentally feasible implementation of one-way quantum computing in superconducting circuits. First, we show how to efficiently generate  2D cluster states by using high-fidelity controlled-Z gates, which are standard in this platform. We study the implementation of feedforward protocols and apply this to the generation of a universal C-NOT gate. Finally, we propose an alternative initial state with maximal persistence which can be straightforwardly generated in superconducting circuits. We show that the implementation of quantum algorithms with this multipartite state requires $25\%$ less ancillary qubits and measurements than with cluster states, substantially improving the involved scalability aspects.

\section{Cluster state and C-NOT gate generation}
First, we propose a digital generation of cluster states, the quantum resource for measurement-based quantum computing. Our protocol is based on the controlled-Z gate ($C^z$) implemented by Barends \textit{et. al.} \cite{Barends2014,Martinis2014} in Xmon qubits \cite{Barends2013}. The Xmon is a tunable non-linear system, which allows to define qubit states $\ket{1}$ and $\ket{0}$ in its ground and first excited state, respectively, and an auxiliary state $\ket{a}$ in the second excited state. As described in Ref. \cite{Barends2014}, in the implementation of the $C^z$ gate the Xmon qubits are capacitively coupled. Tuning the energy gap between the states $\{\ket{0},\ket{a}\}$ of the first Xmon to match the energy gap between the states $\{\ket{1},\ket{0}\}$ of the adjacent Xmon it is possible to activate an exchange evolution  between them that is given by $U=e^{-\frac{i}{\hbar} Vt}$, with $V =\hbar g(\ket{0}\ket{0}\bra{1}\bra{a} + \textrm{H.c.}) $ The $C^z$ gate is realized for $gt = \pi$, that allows to implement the conditional change $\ket{0}\ket{0}\rightarrow-\ket{0}\ket{0}$. 

The cluster state $\ket{C_N}$ of a lattice of $N$ particles is written as $\ket{C_N}=\mathcal{U}\bigotimes_j \ket{+_j}$, where the operator $\mathcal{U}$ is given by the time evolution operator $\mathcal{U}=e^{-iH\pi}$ \cite{Briegel2001}, with
\begin{eqnarray}
H=\sum_{\langle j,k \rangle,k>j}&&\Bigg(\frac{1+\sigma^z_j}{2}\Bigg)\Bigg(\frac{1-\sigma^z_k}{2}\Bigg),\quad\ket{+_j}=\frac{\ket{0_j}+\ket{1_j}}{\sqrt{2}},
\label{Hamiltonian}
\end{eqnarray}
with $\ket{0_j}$ and $\ket{1_j}$ being the eigenstates of $\sigma^z_j$ with eigenvalue $1$ and $-1$ respectively, $\langle j,k \rangle$ denotes that the sites $j$ and $k$ are nearest-neighbour. The effect of $\mathcal{U}$ over each pair of adjacent qubits is a controlled-Z gate, with $\ket{0_j1_k}\rightarrow-\ket{0_j1_k}$, where $j$ is the control and $k$ the target qubit. Therefore, we can write a cluster state as \cite{Briegel2001}
\begin{equation}
\ket{C_N}=\frac{1}{2^{N/2}}\bigotimes_{\ell=1}^{N-1}\Bigg[\ket{0_{\ell}}\Big(\prod_{j}\sigma^z_{\langle\ell_j\rangle}\Big)+\ket{1_{\ell}}\Bigg],
\label{ClusterChain}
\end{equation}
where $\langle\ell_j\rangle$ refers to the $j$th nearest neighbour of $\ell$. 
\begin{figure}[b]
\centering
\includegraphics[width=0.6\linewidth]{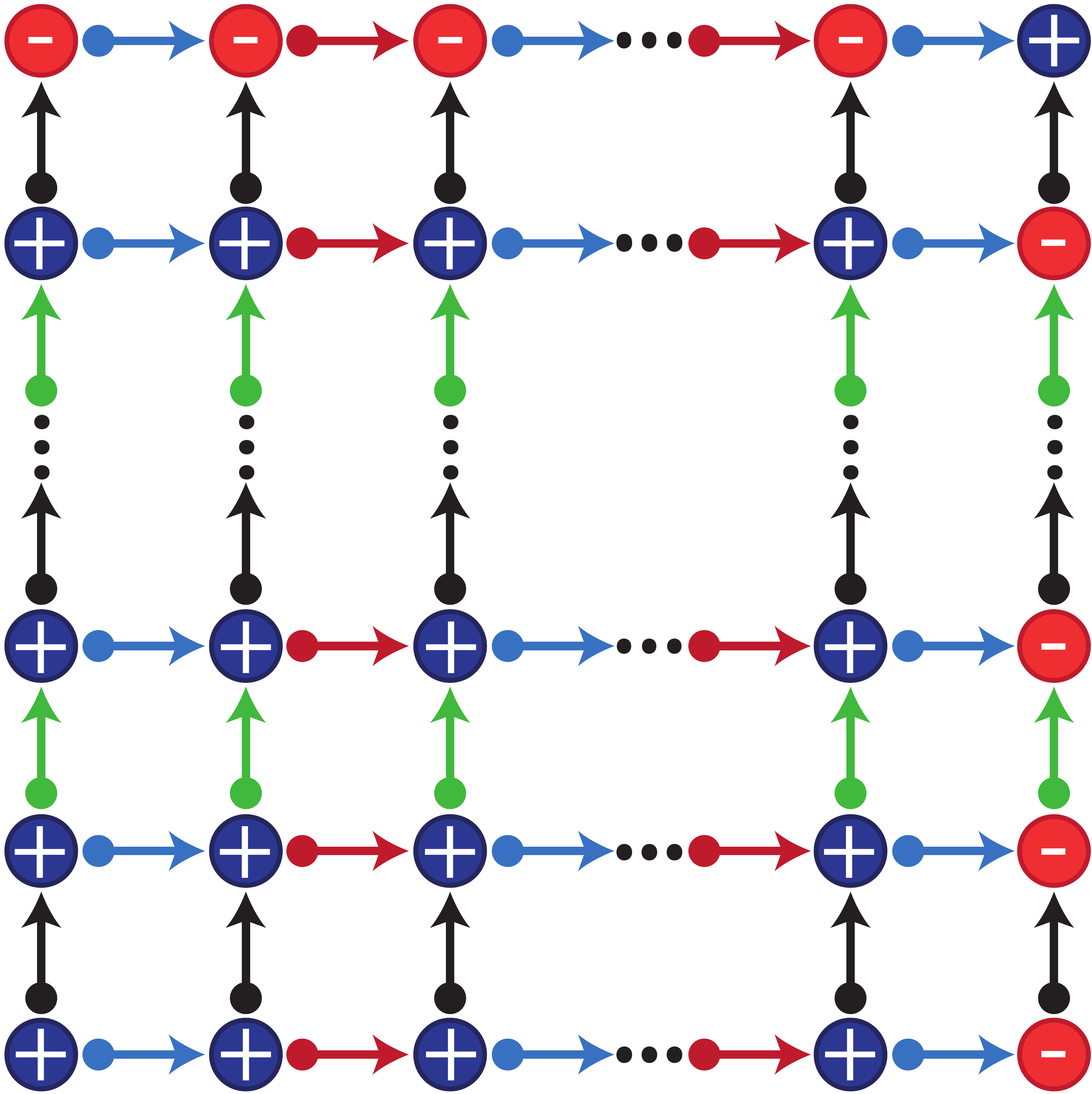}
\caption{Square lattice for the generation of a cluster state. Each link between sites indicates a $\mathcal{C}_z$ gate, where the dot indicates the control site and the arrow the target site. The color of the links indicates the gates that can be performed simultaneously, to perform the blue, red, black and green interactions it requires four steps overall. The sites marked with ``$+$'' are initialized in the $\ket{+}$ state, an the sites marked with ``$-$'' in the $\ket{-}$ state.}
\label{Lattice}
\end{figure}
This effect can be simulated using a combination of $C^z$ gates over Xmon qubits. As described in Ref. \cite{Barends2014}, this gate changes the state $\ket{0_j}\ket{0_k}$ into $-\ket{0_j}\ket{0_K}$, acting as the identity on the remaining states, where $\ket{0}$ and $\ket{1}$ stand for the excited and ground states of the Xmon qubit, respectively. Then, the gate $C^z$ performs a $-\sigma_{z}$ on the target when the control qubit is in the state $\ket{0}$. This  differs slightly from the action of the  $\mathcal{U}$ gate that activates a $\sigma_{z}$ gate on the target. Therefore, to obtain the state given by Eq.~(\ref{ClusterChain}), we change the initial state as follows, the sites that act as a control for an even number of gates are initialized in the $\ket{+}$ state, and, the sites that act as a control for odd number of gates are initialized in the state $\ket{-}=(-\ket{0}+\ket{1})/\sqrt{2}$. In Fig.~\ref{Lattice}, we show the initial state for the square lattice.

As an example we write, the cluster state for a $h\times l$ square lattice. For this case, we define a convenient notation and denote each point in the lattice by a vector $\vec{p}=(j,k)$, where $j$ and $k$ denote the $j$th column and $k$th row respectively. In Fig.~\ref{Lattice} the site $(1,1)$ represent the left-down-corner. We also define $\hat{\imath}=(1,0)$ and $\hat{\jmath}=(0,1)$, and obtain 
\begin{eqnarray}
\ket{C_N^{hl}}=&&\Bigg[\prod_{\vec{p}\in\{\mathcal{L}+\mathcal{L}_x\}}\mathcal{C}^z_{(\vec{p},\vec{p}+\hat{\imath})}\Bigg]\Bigg[\prod_{\vec{p}\in\{\mathcal{L}+\mathcal{L}_y\}}\mathcal{C}^z_{(\vec{p}_x,\vec{p}_x+\hat{\jmath})}\Bigg]\nonumber\\
&&\Bigg[\bigotimes_{\vec{p}\in\mathcal{L}}\ket{+_{\vec{p}}}\Bigg]\Bigg[\bigotimes_{\vec{p}\in \mathcal{L}_x}\ket{-_{\vec{p}}}\Bigg]\Bigg[\bigotimes_{\vec{p}\in \mathcal{L}_y}\ket{-_{\vec{p}}}\Bigg]\otimes\ket{+_{(h,l)}},
\label{lattice1}
\end{eqnarray}
where $\mathcal{L}=\{(j,k)\}$, $\mathcal{L}_x=\{(j,l)\}$, $\mathcal{L}_y=\{h,k\}$, with $j=\{1,2,\dots,h-1\}$ and $k=\{1,2,\dots,l-1\}$. In Fig. \ref{Lattice}, $\mathcal{L}$ contains all blue points, except the right-up-corner, $\mathcal{L}_x$ contains the red points in the right boundary, and $\mathcal{L}_y$ the red points in the upper boundary. 
	
The effect of $C_{j,k}^z$ can be thought of as a time evolution given by $e^{-i\bar{H}_{j,k}\pi}$, with $\bar{H}_{j,k}=(1+\sigma^z_j)(1+\sigma^z_k)/4$, and defining $\mathcal{H}=\sum_{\langle j,k\rangle}\bar{H}_{j,k}$, we rewrite Eq. (\ref{lattice1}) as
\begin{eqnarray}
\ket{C_N^{hl}}=&&e^{-i\mathcal{H}\pi}\Bigg[\bigotimes_{\vec{p}\in\mathcal{L}}\ket{+_{\vec{p}}}\bigotimes_{\vec{p}\in\mathcal{L}_x}\ket{-_{\vec{p}}}\bigotimes_{\vec{p}\in\mathcal{L}_y}\ket{-_{\vec{p}}}\Bigg]\otimes\ket{+_{(h,l)}}.
\label{lattice2}
\end{eqnarray}
Although all components $\bar{H}_{j,k}$ of the Hamiltonian commute, the physical implementation of $\mathcal{C}^z$ prohibits the simultaneous realization of gates that share qubits \cite{Martinis2014}. Then, we separate the Hamiltonian $\mathcal{H}$ in Eq. (\ref{lattice2}) into terms corresponding to the gates that can be performed at the same time and write  $\mathcal{H}=\mathcal{H}^x_1+\mathcal{H}^x_2+\mathcal{H}^y_1+\mathcal{H}^y_2$, where
\begin{eqnarray}
&&\mathcal{H}^x_{\ell}=\sum_{\vec{p}\in A^x_{\ell}} (1+\sigma^{z}_{\vec{p}})(1+\sigma^{z}_{\vec{p}+\hat{\imath}}),\nonumber\\
&&\mathcal{H}^y_{\ell}=\sum_{\vec{p}\in A^y_{\ell}} (1+\sigma^{z}_{\vec{p}})(1+\sigma^{z}_{\vec{p}+\hat{\jmath}}),
\end{eqnarray}
with $\ell=\{1,2\}$.  $A^x_{\ell}$ and $A^y_{\ell}$ correspond to the set of points of the form $(2n_{x}+\ell,k)$ and $(j,2n_{y}+\ell)$, respectively, where $k=\{1,2,\dots,l\}$, $j=\{1,2,\dots,h\}$, $n_x=\{0,1,\dots,h/2-2\}$ and $n_y=\{0,1,\dots,l/2-2\}$. Therefore, we can simulate the evolution $e^{-i\mathcal{H}\pi}$ in $4$ steps. In Fig. (\ref{Lattice}), the controlled-Z gates that produce the evolutions $e^{-i\mathcal{H}_1^x \pi}$, $e^{-i\mathcal{H}_2^x\pi}$, $e^{-i\mathcal{H}_1^y\pi}$ and $e^{-i\mathcal{H}_2^y\pi}$ are represented by blue, red, black and green links respectively. As all Hamiltonians $\mathcal{H}_j^{\alpha}$ commute, no digital error is committed. Finally, the elapsed time reported for the $\mathcal{C}^z$ gate is around $t_{CZ}\approx0.05[\mu s]$ with average fidelity $99.5\%$ \cite{Barends2014}, then for a $4\times4$ lattice, the protocol can generate a cluster state with average fidelity of $88\%$ in $0.2[\mu s]$.
	
\begin{figure}[t]
\centering
\includegraphics[width=0.7\linewidth]{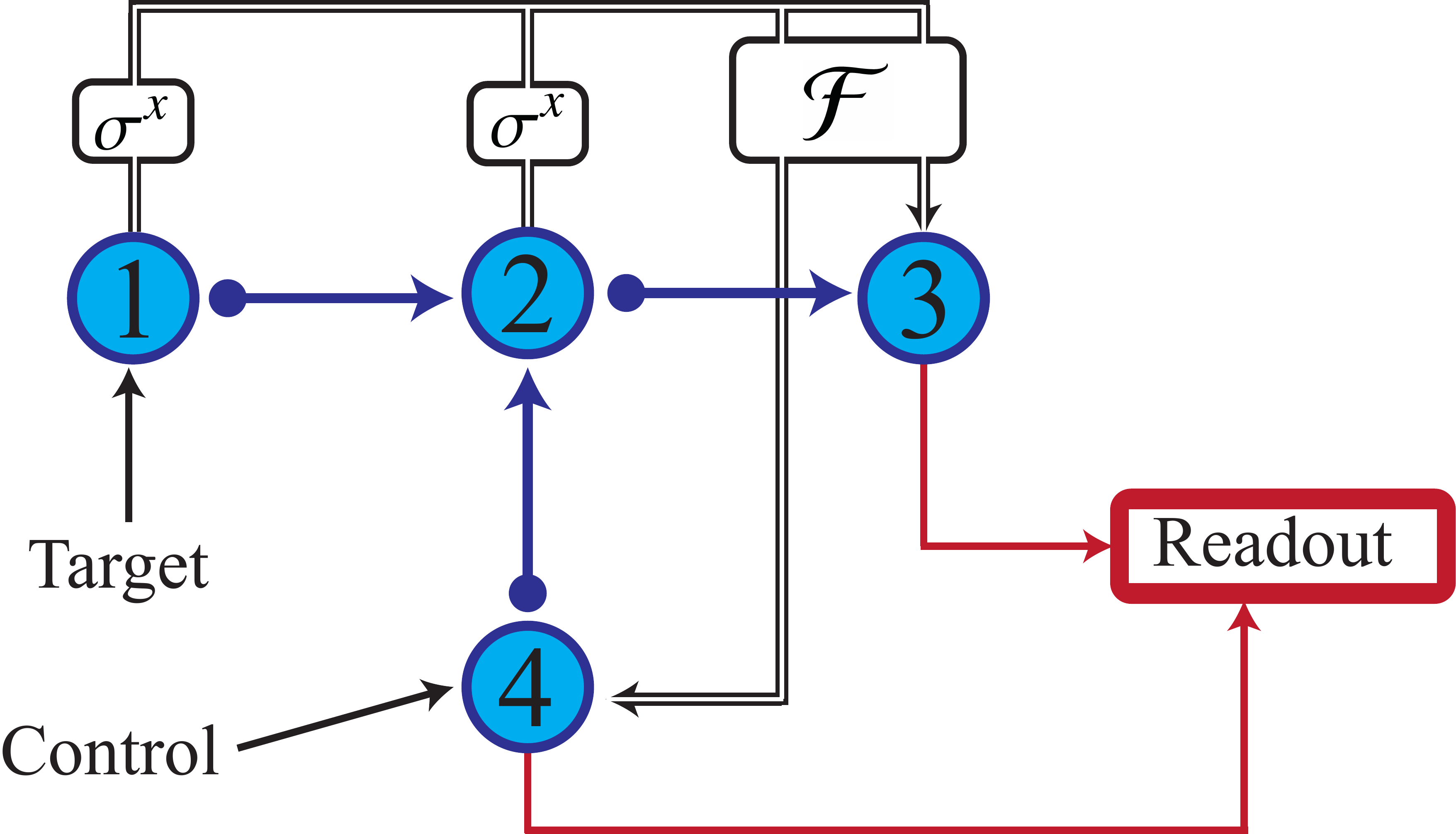}
\caption{Schematic protocol for the C-NOT gate. Qubits $1$ and $4$ are the target and control qubit, respectively, and the sites $2$ and $3$ are initialized in the states $\ket{-}$ and $\ket{+}$, respectively. The arrows indicate the order in which the interaction is considered, where the circles are attached to the control qubits for the CZ gate, and the arrow end to the target. Finally we measure qubits $1$ and $2$, and with this information we perform a feedback over the readout qubits ($3$ and $4$).}
\label{Cnot}
\end{figure}

Consider now, as a specific example of a two-qubit gate, the implementation of the C-NOT. Pedagogically, we will use the previous developed gate as in Ref.~\cite{Raussendorf2001}. Let us use the protocol of Fig.~\ref{Cnot}, where qubit $1$ is the target, qubit $4$ is the control, and the sites $2$ and $3$ are initialized in the states $\ket{-}$ and $\ket{+}$, respectively. The initial state reads
\begin{equation}
\ket{\Phi_o}=\ket{i_1}_z\ket{j_4}_z\ket{1_2}_x\ket{0_3}_x,
\end{equation}
where the sub-index $z$ and $x$ indicate that we use the eigenbasis of $\sigma^z$ and $\sigma^x$, respectively. Now, we perform the entangling gate $U=\mathcal{C}^z_{(1,2)}\mathcal{C}^z_{(4,2)}\mathcal{C}^z_{(2,3)}$, as represented by the arrow links in Fig. (\ref{Cnot}), thus generating the initial resource state for the implementation of the C-NOT gate. Afterwards we measure the sites $1$ and $2$ in the $\sigma^x$ basis with a respective feedforward over the sites $3$ and $4$, obtaining the state
\begin{equation}
\ket{\Phi_m}=\mathcal{F}_{3,4}\ket{l_1}_z\ket{m_2}_z\ket{(i\oplus j)_3}_z\ket{j_4}_z,
\end{equation}
where $\mathcal{F}_{3,4}=\Big(\sigma^z_3\sigma^z_4\Big)^{l}\Big(\sigma^x_3\Big)^{m}$ is the feedforward operator, $l$ and $m$ are the output of the measurement of qubits $1$ and $2$, respectively. The reported elapsed time for the measurement and digital feedforward process is approximately $2[\mu s]$ \cite{Riste2015}, with an error of $1\%$ for the measurement and $1\%$ for the feedforward process. Then, a rough estimation of the fidelity of this protocol yields $95\%$, in a time smaller than $5[\mu s]$, which is significantly smaller than the average Xmon coherence time~\cite{Barends2013}. The elapsed times and fidelity we mentioned before correspond to an effective estimation taking into consideration gate sequences, feedback delays and readout times which have been experimentally reported \cite{Riste2015}. Based on our estimations, we expect that, with current technology, it could be possible to implement quantum algorithms of up to $5$ two-qubit gates and requiring approximately $20$ qubits with a fidelity lower-bounded roughly by $80\%$.
	
\section{Alternative entangling gate}
\begin{figure}[b]
\centering
\includegraphics[width=0.6\linewidth]{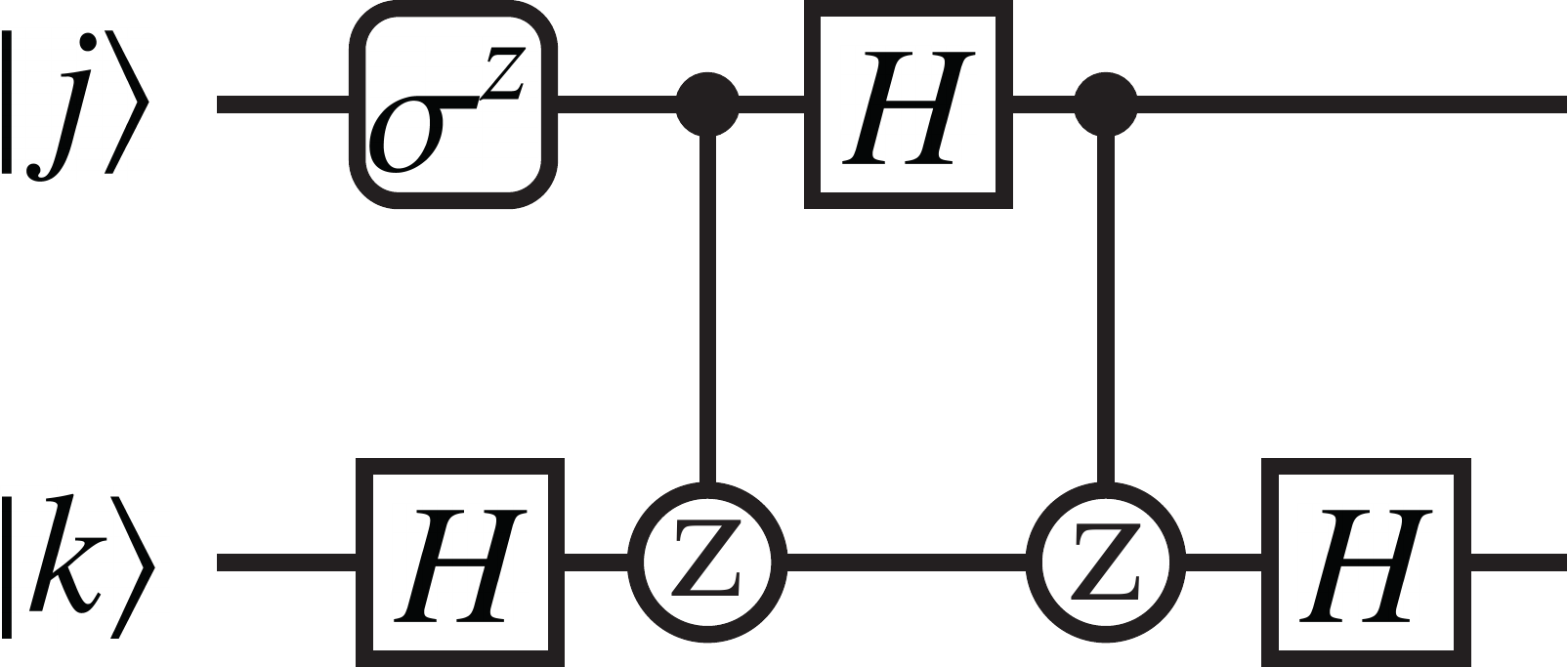}
\caption{Circuit to construct the $U^{\rm Bell}_{j,k}$ gate given in Eq. (\ref{gateB}).}
\label{gate}
\end{figure}

One of the limitations faced by measurement-based quantum computing is the number of qubits required for the implementation of quantum algorithms. In what follows, we propose an alternative entangling gate, which we have termed $U^{\textrm{Bell}}$ and show that it reduces the number of qubits needed for one-way quantum computing. In addition, this gate enables the generation of maximal persistence states, allowing, for instance, the implementation of an efficient C-NOT protocol for one-way quantum computing. We summarize the action of this gate as follows
\begin{eqnarray}
U^{\textrm{Bell}}_{j,k}\ket{0_{j}0_{k}}=\sqrt{\frac{1}{2}}\Big(\ket{0_{j}0_{k}}+\ket{1_{j}1_{k}}\Big) , \nonumber\\
U^{\textrm{Bell}}_{j,k}\ket{0_{j}1_{k}}=\sqrt{\frac{1}{2}}\Big(\ket{0_{j}1_{k}}+\ket{1_{j}0_{k}}\Big) , \nonumber\\
U^{\textrm{Bell}}_{j,k}\ket{1_{j}0_{k}}=\sqrt{\frac{1}{2}}\Big(\ket{1_{j}0_{k}}-\ket{0_{j}1_{k}}\Big) , \nonumber\\
U^{\textrm{Bell}}_{j,k}\ket{1_{j}1_{k}}=\sqrt{\frac{1}{2}}\Big(\ket{1_{j}1_{k}}-\ket{0_{j}0_{k}}\Big) .
\label{gateB}
\end{eqnarray}
Within the Xmon architecture discussed in the previous section, this gate could be implemented using the circuit shown in Fig.~\ref{gate}, which involves one and two-qubit gates among adjacent Xmons. However, it is worthy to mention that the $U^{{\rm Bell}}_{j,k}$ gate could be implemented as the time evolution ${U^{\rm Bell}_{j,k}=e^{-\frac{i}{\hbar} H_{j,k}^{x,y}\tau}}$, with the Hamiltonian ${H_{j,k}^{x,y}=\hbar\xi(J_1\sigma^x_j\sigma^y_k-J_2\sigma^y_j\sigma^x_k})$, where $J_1=5/4$, $J_2=1$, and $\xi\tau=\pi$. It is not clear whether it is possible to implement this Hamiltonian in the Xmon qubit architecture. Nevertheless, the Hamiltonian $H_{j,k}^{x,y}$ can be implemented using a chain of qubit-cavity systems in the ultrastrong coupling regime, with variable coupling, as is shown in a recent theoretical work \cite{AlbarranArriagada2018}, which is summarized in Appendix~\ref{Ap}.

\begin{figure}[b]
\centering
\includegraphics[width=0.6\linewidth]{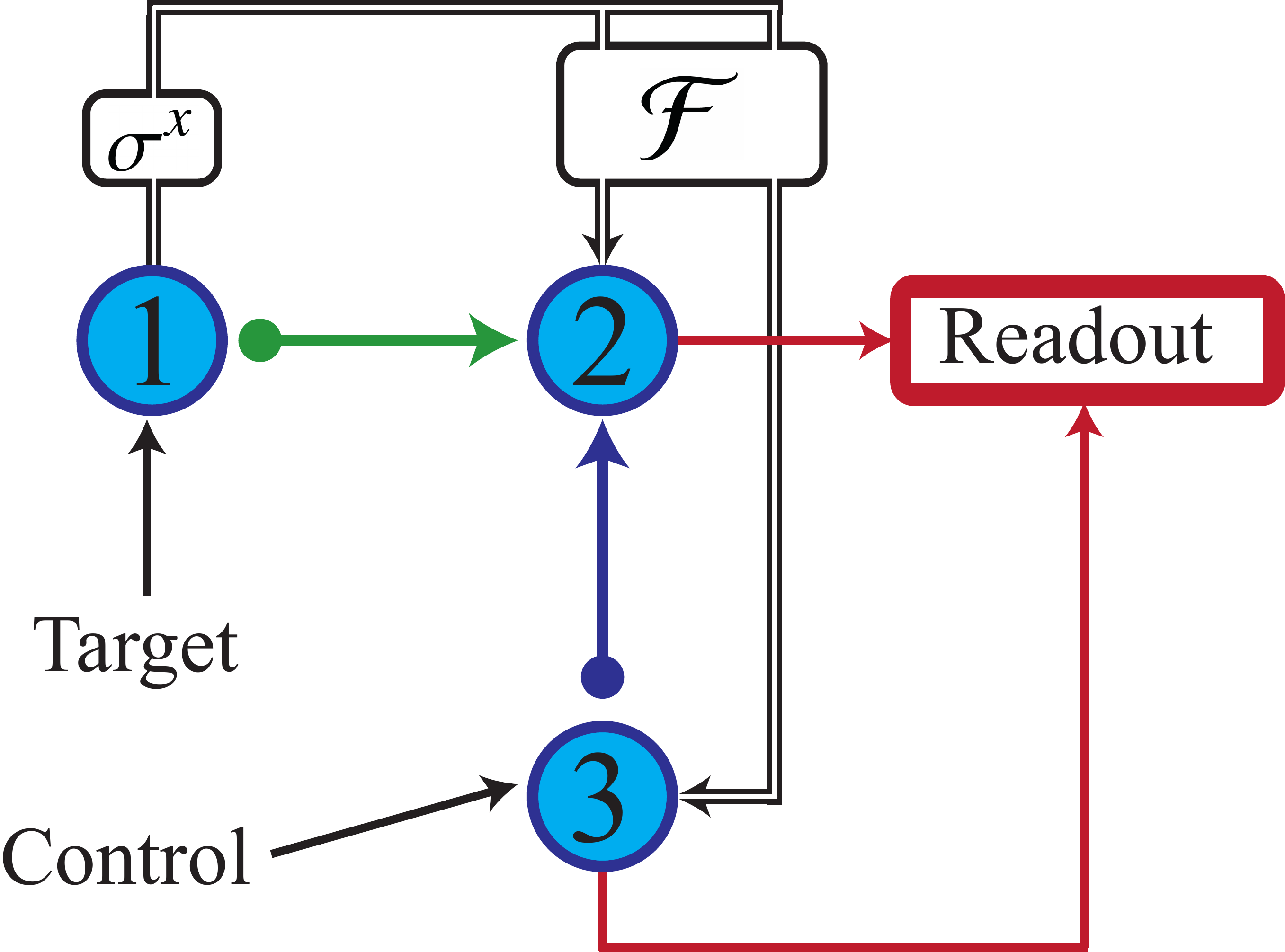}
\caption{Schematic protocol for our efficient C-NOT gate, where qubits $1$ and $3$ are the target and control, respectively, site $2$ is initialized in the state $\ket{0}$. The blue arrow refers to the gate $\mathcal{C}_{3,2}$ and the green arrow for the gate $U^{\rm Bell}_{1,2}$. Finally, we measure qubit $1$, and with this information we perform a feedback over the non-measured qubits ($2$ and $3$).}
\label{Cnot2}
\end{figure}

In what follows, we show that using the $U^{\rm Bell}_{j,k}$ entangling gate it is possible to generate an entangled multiparticle state that plays a similar role as the cluster state as a resource for one-way quantum computing. This state allows to implement a C-NOT gate that only requires $3$ qubits and one measurement, which is an improvement over the $4$ qubits and two measurements needed for the usual cluster state generated by the $\mathcal{C}^z_{(j,k)}$ gate, as was shown in the previous section. The protocol is summarized in Fig.~\ref{Cnot2}. The sites $1$, $2$, and $3$ correspond to target input, target output, and control input qubit, respectively. The initial state is given by
\begin{equation}
\ket{\Psi_o}=\ket{i_1}_z\ket{0_2}_z\ket{j_3}_z .
\end{equation}
Now, we perform the $U^{\rm Bell}_{1,2}$ gate which leads to
\begin{equation}
\ket{\Psi_1}=\sqrt{\frac{1}{2}}\Big(\ket{i_1}_z\ket{0_2}_z+(-1)^i\ket{(i\oplus 1)_1}_z\ket{1_2}_z\Big)\ket{j_3}_z .
\end{equation}
Next, we do the $\mathcal{C}^z_{3,2}$ gate and obtain 
\begin{equation}
\ket{\Psi_2}=\sqrt{\frac{1}{2}}\Big((-1)^{j+1}\ket{i_1}_z\ket{0_2}_z+(-1)^i\ket{(i\oplus 1)_1}_z\ket{1_2}_z\Big)\ket{j_3}_z,
\end{equation}
This state will be the resource that plays a similar role as the cluster state, with the improvement that it only requires a single measurement to implement a C-NOT gate, as will be shown as follows.

Measuring $\sigma^x$ in qubit $1$ yields
\begin{equation}
\ket{\Psi_3}=\sqrt{\frac{1}{2}}\Big((-1)^{j+1+i\cdot s}\ket{s_1}_x\ket{0_2}_z+(-1)^{i+i\cdot s+s}\ket{s_1}_x\ket{1_2}_z\Big)\ket{j_3}_z,
\end{equation}
where $s\in\{0,1\}$, is the outcome of the $\sigma^x$ measurement. Then, we perform a Hadamard gate over qubit $2$, obtaining
\begin{eqnarray}
\ket{\Psi_4}=&&\frac{(-1)^{i\cdot s}}{2}\ket{s_1}_x\bigg[\Big((-1)^{j+1}+(-1)^{i+s}\Big)\ket{0_2}_z\nonumber\\
&&+\Big((-1)^{j+1}-(-1)^{i+s}\Big)\ket{1_2}_z\bigg]\ket{j_3}_z .
\end{eqnarray}
If $s=0$, we obtain $\ket{\Psi_4}=(-1)^{j+1}\ket{0_1}_x\ket{j_3}_z\ket{(i\oplus j\oplus 1)_2}_z$, and if $s=1$, we obtain $\ket{\Psi_4}=(-1)^{i+j+1}\ket{1_1}_x\ket{j_3}_z\ket{(i\oplus j)_2}_z$. Then, when $s=0$ we need to activate the operator $-\sigma^x_2\sigma^z_3$, and when $s=1$ the operator $-\sigma^z_2$ in order to recover the usual control NOT gate, where qubit $2$ becomes the target qubit. Therefore, using feedforward operator $\mathcal{F}=(-\sigma^x_2\sigma^z_3)^{s+1}(-\sigma^z_2)^{s}$, we have $\mathcal{F}\ket{\Psi_4}=\ket{s_1}_x\ket{j_3}_z\ket{(i\oplus j)_2}_z$. As $U^{\rm Bell}_{1,2}$ and $\mathcal{C}^z_{3,2}$ do not commute, the order in which the gates are applied is important.
	
As is shown in Fig. \ref{gate}, the gate $U^{\rm Bell}$ can be constructed using three Hadamard gates, two  CZ-gates and a $\sigma_{z}$ gate. A rough estimation of the fidelity yields $98.8\%$ and requires a time of less than $0.3[\mu s]$. As we have seen, the C-NOT requires one $U^{\rm Bell}$ and one $C^z$ gate. It can be done with a fidelity around $96.5\%$ in a time $2.5[\mu s]$. Therefore, the implementation of quantum algorithms using this protocol requires $25\%$ less ancillary qubits, substancially improving scalability. For example, an array of 16 qubits allows for $4$ C-NOT gates with the standard protocol, whereas it is possible to implement $5$ C-NOT gates with our protocol. Furthermore, since the C-NOT and arbitrary qubit rotations form a universal set~\cite{Barenco1995} we can expect that the reduction in necessary resources enabled by the $U^{\rm Bell}$ entangling gate will extend to more complex quantum algorithms since any quantum algorithm can be decomposed into C-NOT gates and single qubit rotations.
	
Also, the $U^{\rm Bell}$ operator allows the generation of maximally persistent and maximally connected (MPMC) states. We start with the initial state $\ket{\phi_o}=\otimes_{\ell=1}^N\ket{0_{\ell}}$, to which we apply the gate $U^{\rm Bell}_{j,k}$ simultaneously over qubit pairs in the sites $(2\ell-1,2\ell)$, followed by the same operation on qubit pairs in sites $(2\ell,2\ell+1)$, obtaining
\begin{eqnarray}
&&\ket{\mathfrak{C}_N}=\Bigg[\prod_{\ell=1}^{N/2-1}U^{\rm Bell}_{2\ell,2\ell+1}\Bigg]\Bigg[\prod_{\ell=1}^{N/2}U^{\rm Bell}_{2\ell-1,2\ell}\Bigg]\bigotimes_{\ell=1}^{N}\ket{0_{\ell}}\nonumber\\
&&=\frac{1}{2^{N/4}}\Bigg[\prod_{\ell=1}^{N/2-1}U^{\textrm{Bell}}_{2\ell,2\ell+1}\Bigg]\bigotimes_{\ell=1}^{N/2}\Big[\ket{0_{2\ell-1}}\ket{0_{2\ell}}+\ket{1_{2\ell-1}}\ket{1_{2\ell}}\Bigg] . \nonumber\\
\end{eqnarray}
As an example, for three and four particles, we have the state
\begin{equation}
\ket{\mathfrak{C}_3}=U^{\rm Bell}_{2,3}\ket{\mathfrak{C}_2}\ket{0}=\frac{1}{\sqrt{2}}\Big(\ket{\mathfrak{C}_2}\ket{0}+\ket{\mathfrak{C}_2^{\perp}}\ket{1}\Big),
\label{C3}
\end{equation}
where $\ket{\mathfrak{C}_2}=(\ket{0}\ket{0}+\ket{1}\ket{1})/\sqrt{2}$ and $\ket{\mathfrak{C}_2^{\perp}}=(\ket{0}\ket{1}-\ket{1}\ket{0})/\sqrt{2}$ are orthogonal cluster states of two particles. And we also have
\begin{equation}
\ket{\mathfrak{C}_4}=U^{\rm Bell}_{2,3}\ket{\mathfrak{C}_2}\ket{\mathfrak{C}_2}=\frac{1}{\sqrt{2}}\Big(\ket{\mathfrak{C}_3}\ket{0}+\ket{\mathfrak{C}_3^{\perp}}\ket{1}\Big),
\label{C4}
\end{equation}
where $\ket{\mathfrak{C}_3^{\perp}}=(\ket{\mathfrak{C}_2}\ket{1}+\ket{\mathfrak{C}_2^{\perp}}\ket{0})/\sqrt{2}$, is a MPMC-state of $3$ particles orthogonal to $\ket{\mathfrak{C}_3}$. Then, if $\ket{\mathfrak{C}_{N-2}}=(\ket{\mathfrak{C}_{N-3}}\ket{0}+\ket{\mathfrak{C}_{N-3}^{\perp}}\ket{1})/\sqrt{2}$, we have
\begin{eqnarray}
\ket{\mathfrak{C}_N}=&&U^{\textrm{Bell}}_{N-2,N-1}\ket{\mathfrak{C}_{N-2}}\ket{\mathfrak{C}_{2}}\nonumber\\
=&&\frac{1}{\sqrt{2}}\Bigg[\frac{1}{\sqrt{2}}\Bigg(\ket{\mathfrak{C}_{N-2}}\ket{0}+\ket{\mathfrak{C}_{N-2}^{\perp}}\ket{1}\Bigg)\ket{0}\nonumber\\
&&+\frac{1}{\sqrt{2}}\Bigg(\ket{\mathfrak{C}_{N-2}}\ket{1}+\ket{\mathfrak{C}_{N-2}^{\perp}}\ket{0}\Bigg)\ket{1}\Bigg]\nonumber\\
=&&\frac{1}{\sqrt{2}}\Bigg(\ket{\mathfrak{C}_{N-1}}\ket{0}+\ket{\mathfrak{C}_{N-1}^{\perp}}\ket{1}\Bigg)\label{CN} ,
\end{eqnarray}
where $\ket{\mathfrak{C}_{N-1}}=(\ket{\mathfrak{C}_{N-2}}\ket{0}+\ket{\mathfrak{C}_{N-2}^{\perp}}\ket{1})/\sqrt{2}$ and $\ket{\mathfrak{C}_{N-1}^{\perp}}=(\ket{\mathfrak{C}_{N-2}}\ket{1}+\ket{\mathfrak{C}_{N-2}^{\perp}}\ket{0})/\sqrt{2}$ are orthogonal MPMC-states. It is straightforward to show that any two qubits can be projected into a Bell state by measuring all the remaining qubits in the $\sigma^z$ basis, which means that the $\ket{\mathfrak{C}_{N}}$ are maximally connected. Now, to prove maximal entanglement persistence for these states we proceed as follows. Suppose that the state $\ket{\mathfrak{C}_{N-1}}$ has a minimal product state decomposition (MPSD) of $r$ terms and, therefore, an entaglement persistence of $\mathcal{P}=\log_2 (r)$. As $\ket{\mathfrak{C}_{N-1}^{\perp}}$ is equal to $\ket{\mathfrak{C}_{N-1}}$ under local rotations, then $\ket{\mathfrak{C}_{N-1}^{\perp}}$ also has a MPSD of $r$ terms. Thus, we see that the state $\ket{\mathfrak{C}_{N}}$ in Eq.~$(\ref{CN})$ is a decomposition in $2r$ terms. Thus, $\ket{\mathfrak{C}_{N}}$ reads
\begin{equation}
\ket{\mathfrak{C}_N}=\sum_{k=1}^{2r}\Bigg(\lambda_k\bigotimes_{\ell=1}^{N}\ket{\alpha_{\ell}^{(k)}}\Bigg) ,
\label{PD}
\end{equation}
where $\ket{\alpha_{\ell}^{(k)}}$ is the state of particle $\ell$ and $\sum_{k=1}^{2r}\lambda_k^2=1$. If Eq.~(\ref{PD}) is not a MPSD, we can write
\begin{equation}
\ket{\mathfrak{C}_N}=\sum_{k=1}^{2r-2}\Bigg(\lambda_k\bigotimes_{\ell=1}^{N}\ket{\alpha_{\ell}}\Bigg)+\bar{\lambda}\ket{\varphi}\Big(\beta_0\ket{\alpha_a^{2r-1}}+\beta_1\ket{\alpha_a^{2r}}\Big)
\end{equation}
where $\bar{\lambda}^2=\lambda_{2r-1}^2+\lambda_{2r}^2$, $\beta_0=\lambda_{2r-1}/\bar{\lambda}$, $\beta_1=\lambda_{2r}/\bar{\lambda}$, and $\ket{\varphi}$ is a state involving every other particle except particle $a$. This reduction is not possible because $\ket{\mathfrak{C}_{N-1}}$ and $\ket{\mathfrak{C}_{N-1}^{\perp}}$ do not share any terms. Then, $a\ne N$ and, as $\ket{\mathfrak{C}_{N-1}}$ and $\ket{\mathfrak{C}_{N-1}^{\perp}}$ are in their MPSD, we conclude that $a\ne\{1,2,\ldots N-1\}$. Therefore, Eq. (\ref{PD}) is the MPSD with $2r$ terms and persistence $\mathcal{P}=\log_2 (2r)$. As the MPSD of $\ket{\mathfrak{C}_2}$ is $r=2$, for $\ket{\mathfrak{C}_N}$ it is $r=2^{N-1}$, and its persistence $\mathcal{P}=N-1$ is maximal.

\section{Conclusions}
We have shown that current technology in superconducting circuits enables the consideration of measurement-based quantum computing algorithms, avoiding operational time problems which affect scalability in other quantum computing platforms. Initially, we generated a two-dimensional $N\times N$ cluster state by making use of $2N(N-1)$ experimentally available controlled-phase gates. A quantum algorithm is generated by using single qubit measurements and coherent feed-forward. In particular, we applied this approach to the case $N=2$ to implement the universal C-NOT gate with an estimated fidelity lower-bounded by $95\%$. Rough calculations, based on reported experimental parameters, allow us to estimate that we can perform algorithms above $20$ qubits. Additionally, we propose an alternative and experimentally feasible entangling gate, which reduces the number of required qubits in a $25\%$ per implemented C-NOT gate. This fact reduces the number of measurements and feed-forward processes, which improves the fidelity of the protocol. Therefore, the perspective towards scalability and quantum supremacy is open with steady current improvements in mid-sized superconducting quantum platforms.

F.A.-A. acknowledges support from CONICYT Doctorado Nacional 21140432 and Direcci\'on de Postgrado USACH, G.A.B acknowledges support from CONICYT Doctorado Nacional 21140587 and Direcci\'on de Postgrado USACH, G.R. acknowledges funding from FONDECYT under grant No. 1150653, J.C.R.  thanks FONDECYT for support under grant No. 1140194, L.L. acknowledges support from Ram\'on y Cajal Grant RYC-2012-11391, while L.L. M.S. and E.S. are grateful for the funding of Spanish MINECO/FEDER FIS2015-69983-P and Basque Government IT986-16.  

\appendix
\section{Possible implementation of $U^{\textrm{Bell}}$ gate}\label{Ap}
A recent theoretical work \cite{AlbarranArriagada2018} has proposed a system in which the $U^{\textrm{Bell}}$ gate could be implemented. This system consists of a chain of qubit-cavity systems in the ultrastrong coupling regime named quantum Rabi systems (QRSs). In this proposal each QRS is coupled to its nearest-neighbour trough grounded superconducting quantum interference devices (SQUIDs). This chain is described by
\begin{eqnarray}
&&H=\sum_{\ell=1}^{N}\bigg[H^{\textrm{QRS}}_{\ell}+\bigg(P_{\ell}^{\ell,\ell+1}+P_{\ell}^{\ell-1,\ell}\bigg)(a^{\dagger}_{\ell}+a_{\ell})^2\bigg]\nonumber\\
&&-\sum_{\ell=1}^{N-1}\bigg[2\sqrt{P_{\ell}^{\ell,\ell+1} P_{\ell+1}^{\ell,\ell+1}}(a_{\ell}^{\dagger}+a_{\ell})(a_{\ell+1}^{\dagger}+a_{\ell+1})\bigg]\nonumber\\
&&+\sum_{\ell=1}^{N}\bigg[\bigg(Q_{\ell}^{\ell,\ell+1}\bar{\Phi}_{\ell,\ell+1}(t)+Q_{\ell}^{\ell,\ell-1}\bar{\Phi}_{\ell,\ell-1}(t)\bigg)(a^{\dagger}_{\ell}+a_{\ell})^2\bigg]\nonumber\\
&&-\sum_{\ell=1}^{N-1}\bigg[2\sqrt{Q_{\ell}^{\ell,\ell+1} Q_{\ell+1}^{\ell,\ell+1}}\bar{\Phi}_{\ell,\ell+1}(t)(a_{\ell}^{\dagger}+a_{\ell})(a_{\ell+1}^{\dagger}+a_{\ell+1})\bigg],\qquad
\label{Eq.A1}
\end{eqnarray}
where $P_{\ell}^{\ell,\ell+1}$ and $Q_{\ell}^{\ell,\ell+1}$ are time independent constants that depend on the characteristics of each site $\ell$ and coupling between the QRSs $\ell$ and $\ell+1$, $\bar{\Phi}_{\ell,\ell+1}(t)$ are time dependent magnetic signal threading the SQUID that couple the sites $\ell$ and $\ell+1$, $a_{\ell}$ and $a^{\dagger}_{\ell}$ are the annihilation and creation operator of the cavity corresponding to the $\ell$th $QRS$ respectively. Finally $H^{\textrm{QRS}}_{\ell}$ is the Hamiltonian of the $\ell$th QRS given by
\begin{equation}
H^{\textrm{QRS}}_{\ell}=\frac{\hbar \omega^{q}_{\ell}}{2}\sigma_{\ell}^{z}+\hbar\omega^r_{\ell}a_{\ell}^{\dagger}a_{\ell}+\hbar g_{\ell}\sigma_{\ell}^x\left(a_{\ell}^{\dagger}+a_{\ell}\right),  
\label{Eq.A2}
\end{equation}
with $\omega_{\ell}^q$, $\omega_{\ell}^r$ and $g_{\ell}$ are the qubit frequency, cavity frequency, and coupling strength belonging to the $\ell$ site respectively. Also, $\sigma_{\ell}^k$ is the k-Pauli matrix of the qubit in the $\ell$th QRS. We define the QRS basis by $\{\ket{j}_{\ell}\}$ by $H_{\ell}^{QRS}\ket{j}_{\ell}=\lambda_{j}^{\ell}\ket{j}_{\ell}$. As the QRS has an anharmonic spectrum, we write the cavity operators in the QRS basis and truncate to the first excited state in order to obtain an effective two-level-system per site, also we can consider that $P_{\ell}^{\ell,\ell+1}=P_{\ell}$ and $Q_{\ell}^{\ell,\ell+1}=Q_{\ell}$ only depend on the site, and the magnetic flux $\bar{\Phi}_{\ell,\ell+1}(t)=\bar{\Phi}(t)$ is the same for all SQUIDs, then, we obtain for adjacent QRSs
\begin{eqnarray}
H_{(2)}=&&\sum_{\ell=1}^{2}\Big[(\lambda^{\ell}_{0}+2P_{\ell}z^{\ell}_{0,0})\ket{0}_{\ell}\bra{0}+(\lambda^{\ell}_{1}+2P_{\ell}z^{\ell}_{1,1})\ket{1}_{\ell}\bra{1}\Big]\nonumber\\
&&-2\Big[\sqrt{P_{\ell}P_{\ell+1}}+\sqrt{Q_{\ell}Q_{\ell+1}}\bar{\Phi}(t)\Big]\chi_{0,1}^{\ell}\chi_{0,1}^{\ell+1}\hat{X}^{\ell}\hat{X}^{\ell+1}\nonumber\\
&&+\sum_{\ell=1}^{2}2Q_{\ell}\bar{\Phi}(t)\Big[z^{\ell}_{0,0}\ket{0}_{\ell}\bra{0}+z^{\ell}_{1,1}\ket{1}_{\ell}\bra{1}\Big]
\label{Eq.A3}
\end{eqnarray}
where $\hat{X}^{\ell}=\ket{0}_{\ell}\bra{1}+\ket{1}_{\ell}\bra{0}$, $z^{\ell}_{j,j}={_{\ell}}\bra{j}(a_{\ell}+a^{\dagger}_{\ell})^2\ket{j}_{\ell}$, and $\chi_{0,1}^{\ell}=\chi_{1,0}^{\ell}={_{\ell}}\bra{j}(a_{\ell}+a^{\dagger}_{\ell})\ket{j}_{\ell}$. As the first line in Eq. (\ref{Eq.A3}) are the diagonal time independent part of the Hamiltonian, we define the free Hamiltonian by $H_o=\sum_{\ell=1}^{2}\Big[\eta_{0}^{\ell}\ket{0}_{\ell}\bra{0}+\eta_1^{\ell}\ket{1}_{\ell}\bra{1}\Big]$, with $\eta_{k}^{\ell}=\lambda^{\ell}_{k}+2P_{\ell}z^{\ell}_{k,k}$. We can write the Hamiltonian (\ref{Eq.A3}) in the interaction picture with respect to $H_o$ as
\begin{eqnarray}
H_{(2)}^{\textrm{I}}=&&-2\chi_{0,1}^{\ell}\chi_{0,1}^{\ell+1}\Big(\sqrt{P_{\ell}P_{\ell+1}}+\sqrt{Q_{\ell}Q_{\ell+1}}\bar{\Phi}(t)\Big)\nonumber\\
&&\times \Big(e^{-i\Delta t}\ket{1}_{\ell}\bra{0}\otimes\ket{0}_{\ell+1}\bra{1}+e^{-i\delta t}\ket{1}_{\ell}\bra{0}\otimes\ket{1}_{\ell+1}\bra{0}\nonumber\\
&&+e^{i\Delta t}\ket{0}_{\ell}\bra{1}\otimes\ket{1}_{\ell+1}\bra{0}+e^{i\delta t}\ket{0}_{\ell}\bra{1}\otimes\ket{0}_{\ell+1}\bra{1}\Big)\nonumber\\
&&+\sum_{\ell=1}^{2}2Q_{\ell}\bar{\Phi}(t)\Big[z^{\ell}_{0,0}\ket{0}_{\ell}\bra{0}+z^{\ell}_{1,1}\ket{1}_{\ell}\bra{1}\Big].
\label{Eq.A4}
\end{eqnarray}
with $\Delta=(\eta_0^{\ell}-\eta_1^{\ell})-(\eta_0^{\ell+1}-\eta_1^{\ell+1})$, and $\delta=(\eta_0^{\ell}-\eta_1^{\ell})+(\eta_0^{\ell+1}-\eta_1^{\ell+1})$
Finally, as $\bar{\Phi}(t)$ is an external magnetic flux, we can choose $\bar{\Phi}(t)=\gamma_{+}\cos(\Delta t-\pi/2)+\gamma_{-}\cos(\delta t+\pi/2)$, and defining $\gamma_{\pm}=(J_1\pm J_2)/(\chi_{0,1}^{\ell}\chi_{0,1}^{\ell+1}\sqrt{Q_{\ell}Q_{\ell+1}})$, we can approximate Eq. (\ref{Eq.A4}) using the rotating wave approximation (RWA) with respect the eigen-energies of $H_o$ as
\begin{eqnarray}
H_{(2)}^{\textrm{I}}\approx&& i(J_1+J_2)(\ket{0_{\ell}1_{\ell+1}}\bra{1_{\ell}0_{\ell+1}}-\ket{1_{\ell}0_{\ell+1}}\bra{0_{\ell}1_{\ell+1}})\nonumber\\
&&+i(J_1-J_2)(\ket{1_{\ell}1_{\ell+1}}\bra{0_{\ell}0_{\ell+1}}-\ket{0_{\ell}0_{\ell+1}}\bra{1_{\ell}1_{\ell+1}})\nonumber\\
=&&J_1\sigma_{\ell}^x\sigma_{\ell+1}^y-J_2\sigma_{\ell}^y\sigma_{\ell+1}^x
\label{Eq.A5}
\end{eqnarray}
for $J_1$ and $J_2$ small enough to ensure the RWA. The time evolution of the Hamiltonian (\ref{Eq.A5}) would result in the $U^{\textrm{Bell}}_{\ell,\ell+1}$ gate as indicated in the main text.

More details can be found in reference \cite{AlbarranArriagada2018}.

\end{document}